\documentclass[pdftex,a4paper,12pt]{article}

\usepackage{amssymb,amsmath,amsfonts,nomencl,mathrsfs}
\usepackage{amsthm}  
\usepackage[arrow, matrix, curve]{xy}
\usepackage[latin1]{inputenc}
\newcommand{\IC}{\mathbb{C}}
\newcommand{\IR}{\mathbb{R}}

\newcommand{\A}{\mathscr{A}}

\newcommand{\IFF}{\mathscr{F}}

\newcommand{\IN}{\mathbb{N}}

\newcommand*{\longhookrightarrow}%
               {\ensuremath{\lhook\joinrel\relbar\joinrel\rightarrow}}
\newcommand{\pa}{\slash\slash}

\newcommand{\Id}{{\rm d}}

\newcommand{\sm}{\sim_b}

\theoremstyle{plain}            
\newtheorem{theorem}{theorem}[section]

\newtheorem{Theorem}[theorem]{Theorem}

\theoremstyle{definition}       
\newtheorem{Definition}[theorem]{Definition}

\begin{document}

\begin{titlepage}


 \title{ Recent probabilistic results on covariant Schr\"odinger operators on infinite weighted graphs}

   \author{
Batu G\"uneysu\footnote{Institut f\"ur Mathematik, Humboldt-Universit\"at zu Berlin, Germany. E-mail: gueneysu@math.hu-berlin.de}, Ognjen Milatovic\footnote{Department of Mathematics and Statistics, University of North Florida, Jacksonville, FL 32224, USA. E-mail: omilatov@unf.edu} }




\end{titlepage}

\maketitle

\begin{abstract}
We review recent probabilistic results on covariant Schr\"odinger operators on vector bundles over (possibly locally infinite) weighted graphs, and explain applications like semiclassical limits. We also clarify the relationship between these results and their formal analogues on smooth (possibly noncompact) Riemannian manifolds.
\end{abstract}

\setcounter{page}{1}

\section{A review of covariant Schr\"odinger operators on smooth Riemannian manifolds}\label{sed1}

Let us start by taking a look at covariant Schr\"odinger-type operators on Riemannian manifolds: Assume that $E\to M$ is a smooth finite dimensional Hermitian vector bundle over a possibly noncompact smooth Riemannian manifold $M$ without boundary, with a Hermitian covariant derivative $\nabla$ on $E$, which means that $\nabla$ is a complex linear map
\begin{align}
&\nabla: \Gamma_{\mathsf{C}^{\infty}}(M,E)\longrightarrow \Omega^1_{\mathsf{C}^{\infty}}(M,E),\text{ such that for all $\Psi_1,\Psi_2\in \Gamma_{\mathsf{C}^{\infty}}(M,E)$ one has:}\nonumber\\
& \Id (\Psi_1,\Psi_2)= (\nabla \Psi_1,\Psi_2)+(\Psi_1,\nabla\Psi_2 ).
\label{herm}
\end{align}

Then the symmetric nonnegative sesquilinear form
\begin{align}
\tilde{Q}_{\nabla,0}(\Psi_1,\Psi_2)=\int_M \big(\nabla \Psi_1(x),\nabla \Psi_2(x)\big)_x\mathrm{vol}(\Id x),\> \Psi_1,\Psi_2\in \Gamma_{\mathsf{C}^{\infty}_c}(M,E),\label{man}
\end{align}
is closable in the Hilbert space $\Gamma_{\mathsf{L}^2}(M,E)$, as this form corresponds to a non-negative symmetric operator. If $Q_{\nabla,0}:=\overline{\tilde{Q}_{\nabla,0}}$, and if $V:M\to\mathrm{End}(E)$ is a potential (= a measurable self-adjoint section in $\mathrm{End}(E)\to M$) with $V=V^{+}-V^-$ for some potentials $V^{\pm}\geq 0$ such that $|V^+|\in \mathsf{L}^1_{\mathrm{loc}}(M)$ and $|V^-|\in\mathscr{K}(M)$ (the Kato class on $M$), then the form $Q_{\nabla,V}:=Q_{\nabla,0}+Q_V$ (where $Q_V$ is the maximally defined form given by $V$), is symmetric, densely defined, closed, and semibounded (from below), and we can consider the corresponding self-adjoint operator $H_{\nabla,V}$. In this full generality, these operators appear as the nonrelativistic Hamilton operators corresponding to atoms in a magnetic field, possibly with the electron\rq{}s spin taken into account \cite{hydro}. A fundamental result in this context is the \emph{Feynman-Kac formula}: Namely, the integral kernel of the underlying covariant Schr\"odinger semigroup $( \mathrm{e}^{-t H_{\nabla,V}} )_{t\geq 0}$ is given \cite{G2} by the well-defined path integral formula
\begin{align}
\mathrm{e}^{-t H_{\nabla,V}}(x,y)=p(t,x,y)\mathbb{E}^x\big[  \A^{\nabla,V}_t \pa^{\nabla,-1}_t \big|\mathbb{X}_t=y\big]\in \mathrm{Hom}(E_y,E_x),\label{fes1}
\end{align}
where the process $\mathbb{X}$ is a (scaled) Brownian motion on $M$ under $\mathbb{P}^x$ with transition density $p(t,x,y)$, where
$$
\pa^{\nabla}_t\in \mathrm{Hom}(E_{\mathbb{X}_0},E_{\mathbb{X}_t})
$$
is the (unitary) stochastic $\nabla$-parallel transport along the Brownian paths, and
\begin{align*}
& \A^{\nabla,V}_t\in \mathrm{End}(E_{\mathbb{X}_0})\text{  is the path ordered exponential } \\
&\mathscr{A}_{t}^{\nabla,V}=\mathbf{1}_{\mathbb{X}_0}+\sum_{n=1}^{\infty}(-1)^n\int_{t\sigma_n}\pa^{\nabla,-1}_{s_1}V(\mathbb{X}_{s_1})\pa^{\nabla}_{s_1}\cdots \pa^{\nabla,-1}_{s_n}V(\mathbb{X}_{s_n})\pa^{\nabla}_{s_n} \,\Id s_1\,\cdots\,\Id s_n,
\end{align*}
with $t\sigma_n\subset \IR^n$ the $t$-scaled standard $n$-simplex. Along with many other applications, the Feynman-Kac formula can be used to prove the following quantum mechanical result: One has \emph{a semiclassical behaviour of the quantum partition function}, in the sense that for all $\beta>0$ one has \cite{Baer,Si05}
\begin{align}
\frac{\mathrm{tr}\left(\mathrm{e}^{-\beta \hbar H_{\nabla,V/\hbar}}\right)}{ (2\pi)^{-\dim(M)}\hbar^{-\dim(M)/2}  \int_M\int_{\mathrm{T}^*_xM} \mathrm{tr}_{E_x}\left(\mathrm{e}^{-\beta (|v|^2+V(x)) }\right) \Id v \  \mathrm{vol}(\Id x) } \xrightarrow[]{\hbar\to 0+}1,\label{dsda}
\end{align}
which is in fact equivalent to
\begin{align}
(2\pi\hbar )^{\dim(M)/2} \mathrm{tr}\left(\mathrm{e}^{-\beta \hbar H_{\nabla,V/\hbar}}\right) \xrightarrow[]{\hbar\to 0+}  \int_M \mathrm{tr}_{E_x}\big(\mathrm{e}^{-\beta V(x) }\big)  \mathrm{vol}(\Id x).\label{fes2}
\end{align}
In (\ref{dsda}), $\Id v$ stands for the Lebesgue measure on $\mathrm{T}^*_xM$, and it should be noted that in general this result requires some additional global Golden-Thompson control\cite{Si05} on $V$, like $$\int_Mp(\beta\hbar,x,x) \mathrm{e}^{-\beta \min \{\text{eigenvalues of $V(x)$}\}} \mathrm{vol}(\Id x)<\infty.$$

The aim of this note is to translate all of the above data into the setting of infinite graphs, and to present recently obtained weighted graph-analogues of these results.

\section{Covariant Schr\"odinger operators on infinite graphs: Recent results}\label{sed2}

Let $(X,b,m)$ be an arbitrary weighted graph, that is, $X$ is a countable set, $b$ is a symmetric function
$$
b: X\times X\longrightarrow  [0,\infty) \text{ which satisfies  $b(x,x)=0$, $\sum_{y\in X} b(x,y)<\infty$ for all $x\in X$,}
$$
 and $m:X\to (0,\infty)$ is an arbitrary function. We shall interpret $b$ as an edge weight function and write $x\sm y$, if $b(x,y)>0$, and $m$ is interpreted as a vertex weight function. In this setting, a \emph{complex vector bundle} $F\to X$ (over the countable set $X$) with $\mathrm{rank}(F)= \nu\in\IN$ is given by a family $F=\bigsqcup_{x\in X}F_x$ of $\nu$-dimensional complex linear spaces, with the corresponding space of sections 
$$
\Gamma(X,F)=\left.\big\{f\right| f:X\to F, f(x)\in F_x\big\},
$$
which is a $\mathsf{C}(X)$-module, with $\mathsf{C}(X)$ the complex algebra of functions on $X$. If additionally each fiber $F_x$ comes equipped with a complex scalar product $(\bullet,\bullet)_x=(\bullet,\bullet)^F_x$, then $F\to X$ is referred to as a \emph{Hermitian vector bundle}, and the norm and operator norm corresponding to $(\bullet,\bullet)_x$ will be denoted with $|\bullet|_x$.

\begin{Definition} Let $F\to X$ be a complex vector bundle with rank $\nu\in\IN$.\\
(i) An assignment $\Phi$ which assigns to any $x\sm y$ an isomorphism of complex vector spaces $\Phi_{x,y}: F_x\to F_y$ is called a \emph{$b$-connection} on $F\to X$, if one has $\Phi_{y,x}=\Phi^{-1}_{x,y}$ for all $x\sm y$. \\
(ii) If $F\to X$ is Hermitian, then a $b$-connection $\Phi$ on $F\to X$ is called \emph{unitary}, if $\Phi_{x,y}^*=\Phi^{-1}_{x,y}$ for all $x\sm y$.
\end{Definition}

\emph{We fix a Hermitian vector bundle $F\to X$ of rank $\nu\in\IN$, with a unitary $b$-connection $\Phi$ defined on it.} These data determine the sesquilinear form
\begin{align*}
\tilde{Q}_{\Phi,0}(\Psi_1,\Psi_2)=&\>\frac{1}{2}\sum_{x\sm y}b(x,y)\big(\Psi_1(x)-\Phi_{y,x} \Psi_1(y),\Psi_2(x)-\Phi_{y,x} \Psi_2(y)\big)_{x}
\end{align*}
in the Hilbert space $\Gamma_{\ell^2_m}(X,F)$ given by the sections $\Psi\in \Gamma(X,F)$ such that
$$
\left\|\Psi\right\|^2_m:=\sum_{x\in X} |\Psi(x)|^2_x m(x)<\infty,
$$
with domain of $\tilde{Q}_{\Phi,0}$ consisting of finitely supported sections $\Gamma_{c}(X,F)$. Clearly, $\tilde{Q}_{\Phi,0}$ is densely defined, symmetric, and nonnegative, and in fact it is closable (although, in contrast to the Riemannian setting, $\tilde{Q}_{\Phi,0}$ need not come from a symmetric operator). Furthermore, we point out that in general $\tilde{Q}_{\Phi,0}$ is not bounded. However, with 
$$
C(b,m):=\sup_{x\in X} \frac{1}{m(x)}\sum_{y\in X} b(x,y)
$$
one always has 
$$
\tilde{Q}_{\Phi,0}(f,f)\leq 2C(b,m) \|f\|^2_m,
$$ and one often has $C(b,m)<\infty$ in applications (cf. \cite{parti} and the references therein for more details on these facts). Let us explain the analogy of $\tilde{Q}_{\Phi,0}$ to the Riemannian case (\ref{man}): Firstly, we have the edge vector bundle
$$
F^b:=\bigsqcup_{(x,y)\in X^b} F^b_{(x,y)}:=\bigsqcup_{(x,y)\in X^b} F_{x}\longrightarrow  X^b:= \{b>0\}\subset X\times X,
$$
which with $\Omega^1(X,F;b):= \Gamma(X^b,F^b)$ carries the canonical Hermitian structure
$$
\big(\alpha_1(x,y),\alpha_2(x,y)\big)_{(x,y)}:=\big(\alpha_1(x,y),\alpha_2(x,y)\big)^F_{x}, \>\>\alpha_1,\alpha_2\in \Omega^1(X,F;b).
$$
Then $\Phi$ induces the complex linear map
\begin{align*}
&\nabla_{\Phi}:  \Gamma(X,F)\longrightarrow \Omega^1(X,F;b),\>\>\nabla_{\Phi} \Psi(x,y):=\Phi_{y,x}\Psi(y)-\Psi(x),
\end{align*}
which satisfies the Leibnitz rule
\begin{align*}
&\nabla_{\Phi} (f\Psi)(x,y)=\Id f(x,y) \Psi(x)+f(y)\nabla_{\Phi}\Psi(x,y),\>\>f\in \mathsf{C}(X), \\
&\text{where }\>\Id: \mathsf{C}(X)\longrightarrow \Omega^1(X;b):= \mathsf{C}(X^b),\>\Id f(x,y):= f(y)-f(x)
\end{align*}
is the covariant derivative corresponding to the identity connection on the trivial complex line bundle $X\times \IC\to X$ over $X$. Furthermore, the unitarity of $\Phi$ implies that $\nabla_{\Phi}$ is Hermitian in the sense that for all $\Psi_1,\Psi_2\in \Gamma(X,F)$ one has
$$
\Id \big(\Psi_1,\Psi_2\big)(x,y)= \big(\nabla_{\Phi} \Psi_1(x,y),\Psi_2(x)\big)_x-\big(\Psi_1(y),\nabla_{\Phi^{-1}}\Psi_2(x,y)\big)_y,
$$
which clearly is a discrete analogue to (\ref{herm}). Finally, for $\Psi_1,\Psi_2\in \Gamma_c(X,F)$ one has
$$
 \tilde{Q}_{\Phi,0}(\Psi_1,\Psi_2)=\>\frac{1}{2}\sum_{x\sm y}b(x,y)\big(\nabla_{\Phi}\Psi_1(x,y),\nabla_{\Phi}\Psi_2(x,y)\big)_{(x,y)},
$$
which is of type (\ref{man}) in the situation of \lq\lq{}unweighted edges\rq\rq{} $b(x,y)\in \{0,1\}$.\\
Let $Q_{\Phi,0}:=\overline{\tilde{Q}_{\Phi,0}}$, and note that the canonical scalar regular Dirichlet form $Q$ on $\ell^2_m(X)$ and its associated operator $H$ arise as special cases of the above construction, upon taking the identity connection on $X\times \IC\to X$. Given a \emph{potential} $V$ on $F\to X$, that is, $V\in\Gamma(X,\mathrm{End}(F))$ is pointwise self-adjoint, we can define a symmetric sesqui-linear form $Q_V$ in $\Gamma_{\ell^2_m}(X,F)$ by
$$
Q_V(\Psi_1,\Psi_2)=\sum_{x\in X} \big( V(x)\Psi_1(x),\Psi_2(x)\big)_{x}m(x),\>\>\mathsf{D}(Q_V)=\Gamma_{\ell^2_m\cap\ell^2_{|V|\cdot m}}(X,F).
$$
Let $\mathscr{K}(Q)\supset \ell^{\infty}(X)$ be the Kato class corresponding to $Q$, that is, $w:X\to\IC$ is in $\mathscr{K}(Q)$, if and only if
$$
\lim_{t \to 0+} \sup_{x\in X}\int^t_0\int_X \mathrm{e}^{-s H}(x,y) |w(y)| m(y)\Id s =0.
$$

\begin{Definition}[and Proposition] $V$ is called \emph{Kato decomposable}, if it admits a decomposition $V=V^{+}-V^{-}$ into potentials $V^{\pm}\geq 0$ such that $|V^-|\in\mathscr{K}(Q)$. In this situation, for any $\hbar >0$ the form $Q_{\Phi,V/\hbar}:=Q_{\Phi,0}+Q_{V/\hbar}$ is densely defined, symmetric, closed and semi-bounded, and the self-adjoint operator corresponding to $Q_{\Phi,V/\hbar}$ will be denoted with $H_{\Phi,V/\hbar}$.
\end{Definition}

\emph{We fix a Kato decomposable potential $V$ on $F\to X$}, where we refer the reader to \cite{MilTu} for essential self-adjointness properties of $H_{\Phi,V}$. Let us now prepare the ingredients for the Feynman-Kac formula: As $Q$ is a regular Dirichlet form on a nice space, we can associate a reversible strong right-Markoff process to it. A convenient version
$$
\mathbb{X}: [0,\tau)\times \Omega\longrightarrow  X, \text{ with lifetime $\tau:\Omega\longrightarrow  [0,\infty]$,}
$$
of this process has been constructed in \cite{GKS-13} and the references therein, on a filtered probability space $(\Omega,\IFF,\IFF_*,\mathbb{P})$. Let $\tau_n:\Omega\to [0,\infty]$, $n\in \IN_0$, denote the $n$-th jump time of $\mathbb{X}$, and let $N(t):\Omega\to \IN_0\cup \{\infty\}$ be its number of jumps until $t\geq 0$. Many path properties of this process have been derived in \cite{GKS-13, gmt}. We shall need in the sequel that $\tau=\sup_n \tau_n$ and
\begin{align}
\mathbb{P}\big(b(\mathbb{X}_{\tau_n},\mathbb{X}_{\tau_{n+1}}) >0\text{ for all $n\in\IN_0$}\big) =1,\>\{N(t)<\infty\}=\{t<\tau\}\>\text{ for all $t\geq 0$.}\label{ds}
\end{align}
In particular, the $\Phi$-parallel transport along the paths of $\mathbb{X}$ is well-defined by
\begin{align*}
&\pa^{\Phi}: [0,\tau)\times \Omega\longrightarrow F \boxtimes F^* =\bigsqcup_{(x,y)\in X\times X}\mathrm{Hom}(F_y,F_x)\\
&\pa^{\Phi}_t:=
\begin{cases}
\mathbf{1}_{\mathbb{X}_0},\> \text{ if $N(t)=0$}\\ \\
 \Phi_{\mathbb{X}_{\tau_{N(t)-1}},\mathbb{X}_{\tau_{N(t)}}}\cdots \Phi_{\mathbb{X}_{\tau_{0}},\mathbb{X}_{\tau_{1}}}\>\text{ else}
\end{cases}\in \mathrm{Hom}(F_{\mathbb{X}_0},F_{\mathbb{X}_t}),
\end{align*}
which gives a pathwise unitary process, and we can also define the process
$$
\mathscr{A}^{\Phi,V}:[0,\tau)\times\Omega\longrightarrow \mathrm{End}(F)
$$
as the path ordered exponential
\begin{align*}
&\mathscr{A}_{t}^{\Phi,V}-\mathbf{1}_{\mathbb{X}_0}\\
&=\sum_{n=1}^{\infty}(-1)^n\int_{t\sigma_n}\pa^{\Phi,-1}_{s_1}V(\mathbb{X}_{s_1})\pa^{\Phi}_{s_1}\cdots \pa^{\Phi,-1}_{s_n}V(\mathbb{X}_{s_n})\pa^{\Phi}_{s_n} \,\Id s_1\,\cdots\,\Id s_n\in\mathrm{End}(F_{\mathbb{X}_0}).
\end{align*}
Generalizing the scalar magnetic situation from\cite{GKS-13}, the following Feynman-Kac formula, which is our discrete analogue of (\ref{fes1}), has been proven in \cite{gmt}:

\begin{Theorem} With $\mathbb{P}^x:=\mathbb{P}(\bullet|\mathbb{X}_0=x)$, the integral kernel
$$
[0,\infty)\times X\times X\ni (t,x,y)\longmapsto \mathrm{e}^{-t H_{\Phi,V}}(x,y)\in \mathrm{Hom}(F_y,F_x)\subset F \boxtimes F^*,\nonumber
$$
is given by 
\begin{align}
\mathrm{e}^{-t H_{\Phi,V}}(x,y)=m(y)^{-1}\mathbb{P}^x(\mathbb{X}_t=y)\mathbb{E}^x\big[  \A^{\Phi,V}_t \pa^{\Phi,-1}_t \big|\mathbb{X}_t=y\big],\label{feds}
\end{align}
in other words, one has the representation
$$
\mathrm{e}^{-t H_{\Phi,V}}\Psi(x)=\sum_{y\in X}\mathrm{e}^{-t H_{\Phi,V}}(x,y) \Psi(y)m(y),\> \>\>t\geq 0,\>\Psi\in\Gamma_{\ell^2_m}(X,F),\> x\in X.
$$
\end{Theorem}

And the following semiclassical limit is our discrete variant of (\ref{fes2}):

\begin{Theorem}\label{main2} Assume that there is a scalar Kato decomposable function $w:X\to\IR$ with $V\geq w$. Then for all $\beta>0$ with $\sum_{x\in X}\mathrm{e}^{-\beta w(x)}<\infty$, one has
\begin{align}
&\mathrm{tr}(\mathrm{e}^{-\beta\hbar H_{\Phi,V/\hbar}}) \leq  \sum_{x\in X}\mathrm{tr}_x(\mathrm{e}^{-\beta V(x)})<\infty, \label{lim2}\\
&\mathrm{tr}(\mathrm{e}^{-\hbar\beta H_{\Phi,V/\hbar}}) \xrightarrow[]{\hbar\to 0+} \sum_{x\in X}\mathrm{tr}_x(\mathrm{e}^{-\beta V(x)}).\label{lim3}
\end{align}
\end{Theorem}

Let us remark here that $(X,b,m)$ is completely arbitrary in these results (in particular, $(X,b)$ may be locally infinite, and we allow $\inf m=0$). Furthermore, if $(X,b)$ does not support a symmetry which is respected by $m$ and $\mathbb{P}^{\bullet}$ appropriately, then the Golden-Thompson bound (\ref{lim2}) does not follow directly from (\ref{feds}), but rather from a combination of (\ref{feds}) for $V=0$ with the abstract operator variant of the Golden-Thompson bound\cite{GKS-13,parti} (and a combination of geometric and functional analytic approximation arguments). The proof of (\ref{lim3}) uses semigroup domination and the corresponding result in the scalar \lq\lq{}nonmagnetic\rq\rq{} situation, which itself makes full use of the path properties of $\mathbb{X}$. Finally, we would like to point out that by combining the above results with the fact that Brownian motion on (noncompact) Riemannian manifolds can be approximated in law by time continuous geodesic random walks (cf. \cite{kuwada}, where even the situation of time-dependent Riemannian metrics is considered), it should be possible to approximate spectral data of the covariant Schr\"odinger operators from Section \ref{sed1} by the discrete ones of this Section. This should be possible under very general assumptions on the underlying data. We refer the reader to \cite{albeverio,eisen} for special cases in the flat Euclidean space.


\begin{thebibliography}{99}

\bibitem{albeverio} S.~Albeverio, R.~H\o egh-Krohn, H.~Holden and T.~Kolsrud, A covariant Feynman--Kac formula for unitary bundles over Euclidean space, in {\em Stochastic Partial Differential Equations and Applications II (Trento, 1988)}, Lecture Notes in Math., Vol.~1390 (Springer, Berlin, 1989), pp.~1--12.

\bibitem{Baer} C.~B\"ar and F.~Pf\"affle, Asymptotic heat kernel expansion in the semi-classical limit, {\em Comm. Math. Phys.} {\bf 294}, 731--744 (2010).


\bibitem{G2} B.~G\"uneysu, On generalized Schr\"odinger semigroups, {\em J. Funct. Anal.} {\bf 262}, 4639--4674 (2012).

\bibitem{hydro} B.~G\"uneysu, Nonrelativistic hydrogen type stability problems on nonparabolic 3-manifolds,
{\em Ann. Henri Poincar\'e} {\bf 13}, 1557--1573 (2012).

\bibitem{parti} B.~G\"uneysu, Semiclassical limits of quantum partition functions on infinite graphs, arXiv:1402.2452.

\bibitem{GKS-13} B.~G\"uneysu, M.~Keller and M.~Schmidt, A Feynman--Kac--It\^o formula for magnetic Schr\"odinger operators on graphs, arXiv:1301.1304v2.

\bibitem{gmt} B.~G\"uneysu, O.~Milatovic and F.~Truc, Generalized Sch\"odinger semigroups on infinite graphs, {\em Potential Analysis.} DOI 10.1007/s11118-013-9381-6 (2013).

\bibitem{kuwada} K.~Kuwada,
Convergence of time-inhomogeneous geodesic random walks and its application to coupling methods, {\em Ann. Probab.} {\bf 40}, 1945--1979 (2012).

\bibitem{MilTu} O.~Milatovic and F.~Truc, Essential self-adjointness of Schr\"odinger operators on vector bundles over infinite graphs, arXiv:1307.1213.

\bibitem{Si05} B.~Simon, {\em Functional Integration and Quantum Physics}, 2nd edn. (AMS Chelsea Publishing, Providence, 2005).


\bibitem{eisen} V.~S.~Varadarajan and D.~Weisbart, Convergence of quantum systems on grids, {\em J. Math. Anal. Appl.} {\bf 336}, 608--624 (2007).

\end{thebibliography}
\end{document}